\documentclass[useAMS,usenatbib]{mn2e}
\usepackage{times}
\usepackage{graphicx}

\def\aap{A\&A}%
\def\aaps{A\&AS}%
\def\aj{AJ}%
\def\apj{ApJ}%
\def\apjs{ApJS}%
\def\apss{Ap\&SS}%

\def\iaucirc{IAU Circ.}%
\def\mnras{MNRAS}%
\def\pasp{PASP}%
\def\ssr{Space Sci.~Rev.}%
\title[Life after eruption -- V]{Life after eruption -- V. 
Spectroscopy of eight candidate old novae with Gemini-South}
\author[C. Tappert et al.]%
{
C. Tappert,$^{1}$\thanks{E-mail: claus.tappert@uv.cl}
N. Vogt,$^{1}$ 
L. Schmidtobreick$^{2}$
and
A. Ederoclite$^{3}$
\footnotemark[1]\thanks{Based on observations with Gemini-South, program
ID GS-2014A-Q-41, and ESO telescopes, proposal number 089.D-0505(A)}\\
$^{1}$Instituto de F\'{\i}sica y Astronom\'{\i}a, Universidad de 
Valpara\'{\i}so, Avda. Gran Breta\~na 1111, 2360102 Valpara\'{\i}so, Chile\\
$^{2}$European Southern Observatory, Alonso de Cordova 3107, 7630355 Santiago, 
Chile\\
$^{3}$Centro de Estudios de F\'{\i}sica del Cosmos de Arag\'on, Plaza San 
Juan 1, Planta 2, Teruel, E44001, Spain\\
}
\begin{document}

\date{Accepted. Received}

\pagerange{\pageref{firstpage}--\pageref{lastpage}} \pubyear{2013}

\maketitle

\label{firstpage}

\begin{abstract}
We present the analysis of photometric and spectroscopic data on eight 
candidates for post-nova systems. Five post-novae, V528 Aql, HS Sge, BS Sgr, 
GR Sgr and V999 Sgr, are successfully recovered. We furthermore identify likely
candidates for the fields of V1301 Aql, V1151 Sgr and V3964 Sgr. The
spectroscopic properties of the confirmed post-novae are briefly discussed.
We find that two of the oldest post-novae in our sample, GR Sgr and V999 Sgr,
contain an optically thick accretion disc, and thus can be suspected to have a 
high mass-transfer rate, contrary to what one would expect from most models. 
HS Sge and V528 Aql show evidence for a (comparatively) high system 
inclination, which makes them attractive targets for time series observations. 
Finally, the presence of particularly strong \mbox{He\sc ii} emission together 
with a small eruption amplitude suggests that BS Sgr is a good candidate for 
an intermediate polar.
\end{abstract}


\section{Introduction}
\defcitealias{tappertetal12-1}{Paper I}%
\defcitealias{tappertetal13-1}{Paper II}%
\defcitealias{tappertetal13-2}{Paper III}%
\defcitealias{tappertetal14-1}{Paper IV}%

In a cataclysmic variable (CV) a late-type main-sequence (secondary) star 
transfers mass via Roche lobe overflow to a white dwarf (primary), the latter 
being the more massive component. Depending on the strength of the magnetic 
field of the primary, the matter is accreted either via a disc and an inner
boundary layer or along the magnetic field lines directly on the magnetic
pole(s) of the white dwarf. Intermediate configurations, where only the 
inner part of the disc is disrupted by the magnetic field, are also known.
For comprehensive information on CVs see the books by \citet{warner95-1} and 
\citet{hellier01-1}.

Once a certain critical mass has accumulated on the surface of the white
dwarf, a thermonuclear explosion is triggered that ejects the accreted
material. 
This phenomenon is known as a (classical) nova eruption and
represents the principal mechanism for mass-loss in CVs. It is not yet
clear whether more or less than the accreted material is ejected, i.e.~%
whether the white dwarfs in CVs gain or lose mass in the course of CV
evolution. For recent reviews on novae see \citet{bode+evans12-1} and
\citet{woudt+ribeiro14-1}.

The underlying binary system is not destroyed in the nova eruption, but instead
recommences mass transfer within a couple of years \citep*{retteretal98-2}.
This has thus to be a recurrent event, with the minimum recurrence time
currently being estimated to $t_\mathrm{rec} > 10^3~\mathrm{yr}$ 
\citep{sharaetal12-3}. 
In contrast, the class of `recurrent novae' includes novae with recurrence 
times $\le$ 100 yr \citep[and even as short as $\sim$1 yr;][]%
{tangetal14-4,darnleyetal14-1}. They form a small and very heterogeneous 
sample of objects
that in many respects are very similar to CVs and in others show significant
differences \citep[e.g.][]{schaefer10-2,anupama13-1}. In this study
we will thus concentrate on classical novae.

From the above, it follows that CVs can be regarded 
as novae in-between eruptions
\citep{vogt82-2,sharaetal86-1}. It is still unclear to which extent the 
physical parameters of the CV -- magnetic field, chemical composition,
the mass-transfer rate $\dot{M}$ -- affect the nova recurrence time,
although $\dot{M}$ surely appears as the most important one 
\citep{townsley+bildsten05-1}. Similarly, the consequence of the nova
eruption for the (short-- and long-term) evolution of CVs is not yet
understood 
\citep*{prialnik+shara86-1,sharaetal86-1,zorotovicetal11-1,wheeler12-2}.

One possibility to address these questions observationally is to compare
the properties of the group of post-novae with those of other CVs. However,
to be statistically significant such endeavour needs samples of sufficient 
size, and the group of post-novae still does not fulfil this condition. To
alleviate this situation we have started a project to identify candidate
post-novae via colour-colour diagrams and to confirm them spectroscopically
\citep[][hereafter Paper I and IV, respectively]{tappertetal12-1,tappertetal14-1},
determine the orbital period for the brighter post-novae 
\citep[][Paper III]{tappertetal13-2} and to study particularly interesting 
systems in more detail \citep[][Paper II]{tappertetal13-1}. 

In this fifth instalment of our
series, we present data on eight candidate post-novae. The targets were 
selected from the \citet{downesetal05-1} catalogue of CVs, where they were 
marked as potential classical novae. In each case, the reported nova 
eruption took place more than 30 yr ago, which usually represents sufficient
time for the ejected material to disperse and loose its dominance on the 
optical spectrum, so that the latter can be used to examine the underlying CV.
The previous identifications of three 
objects, HS Sge, BS Sgr and V999 Sgr, were found to be ambiguous, and they were 
thus defined as targets for $U\!BV\!R$ photometry to select likely 
candidates. Subsequently, long-slit spectroscopy data were obtained for the
selected objects to confirm or reject the nova hypothesis. Another five 
suspected novae, V528 Aql, V1301 Aql, GR Sgr, V1151 Sgr and V3964 Sgr, 
appeared to be sufficiently well identified to include them directly as 
targets for spectroscopy.

In the following, we describe the observations and the employed techniques
for data reduction (Section \ref{obs_sec}), provide short reviews of the
previously available information on the individual targets and present our 
results (Section \ref{results_sec} and respective subsections), briefly
discuss their implications (Section \ref{disc_sec}), and conclude with a 
summary of the most important results (Section \ref{concl_sec}). Finally,
finding charts for the targets are collected in an appendix.

This project is still ongoing, and it is planned to further increase the
sample of confirmed post-novae before discussing their properties in the 
context of CV evolution in a concluding paper.

\section{Observations and reduction}
\label{obs_sec}

\begin{table}
\caption[]{Calibration values of the $U\!BV\!R$ photometry in magnitudes.}
\label{ubvrcalib_tab}
\begin{tabular}{@{}lllll}
\hline\noalign{\smallskip}
Filter & Night & Extinction & Zero point$^a$ & Colour term \\
\hline\noalign{\smallskip}
$U$ & 2012-06-24 & 0.448(61)$^b$ & $-$0.25(15) & 0.230(51) \\
    & 2012-07-18 & 0.448(61)$^b$ & $-$0.208(20) & $-$0.003(21) \\
    & 2012-07-23 & 0.506(77) & 0.16(11) & $-$0.076(38) \\
$B$ & 2012-06-24 & 0.204(33) & $-$2.768(38) & $-$0.125(39) \\
    & 2012-07-18 & 0.190(30) & $-$2.746(30) & $-$0.125(24) \\
    & 2012-07-23 & 0.190(49) & $-$2.765(29) & $-$0.125(22) \\
$V$ & 2012-06-24 & 0.147(13) & $-$3.035(14) & $-$0.076(07) \\
    & 2012-07-18 & 0.111(25) & $-$2.986(25) & $-$0.074(28) \\
    & 2012-07-23 & 0.125(30) & $-$2.996(18) & $-$0.079(19) \\
$R$ & 2012-06-24 & 0.076(19) & $-$3.086(24) & 0.039(44) \\
    & 2012-07-18 & 0.055(16) & $-$3.017(17) & $-$0.015(22) \\
    & 2012-07-23 & 0.049(25) & $-$3.047(18) & 0.070(46) \\
\hline\noalign{\smallskip}
\end{tabular}
\\
$^a$ With respect to a reference value of 25 mag. \\
$^b$ Tabulated value.
\end{table}

\begin{table*}
\label{obslog_tab}
\begin{minipage}{2.0\columnwidth}
\caption[]{Log of observations.}
\setlength{\tabcolsep}{0.15cm}
\begin{tabular}{@{}lllllllll}
\hline\noalign{\smallskip}
Object & R.A. (2000.0) & Dec.~(2000.0) & rms (arcsec) & Date & Filter/Grism 
& $t_\mathrm{exp}$ (s) & mag & Post-nova? \\
\hline\noalign{\smallskip}
V528 Aql  & 19:19:19.08 & $+$00:37:53.3 & 0.13 & 2014-04-04 & B600 (1.5 arcsec)
& 3000                   & 18.7R & Y \\
V1301 Aql & 19:17:55.20 & $+$04:47:18.3 & --   & 2014-05-01 & B600 (1.5 arcsec)
& 5391                   & 18.9R & N \\  
HS Sge    & 19:39:22.11 & $+$18:07:54.6 & 0.37 & 2012-06-24 & $U$/$B$/$V$/$R$
& 1576 / 420 / 160 / 120 & 20.0V & Y \\
          &             &               &      & 2014-05-03 & B600 (1.5 arcsec)
& 5391                   & 19.6R & \\
BS Sgr    & 18:26:46.39 & $-$27:08:21.6 & 0.23 & 2012-07-18 & $U$/$B$/$V$/$R$
& 1576 / 420 / 160 / 120 & 17.9V & Y \\
          &             &               &      & 2014-04-27 & B600 (1.5 arcsec)
& 5392                   & 17.4R & \\
GR Sgr    & 18:22:58.50 & $-$25:34:47.3 & 0.18 & 2014-04-02 & B600 (1.5 arcsec)
& 241                    & 15.7R & Y \\
V999 Sgr  & 18:00:05.70 & $-$27:33:13.8 & 0.32 & 2012-07-23 & $U$/$B$/$V$/$R$
& 1576 / 420 / 160 / 120 & $<$16.5V & Y \\
          &             &               &      & 2014-04-22 & B600 (1.5 arcsec)
& 450                    & 16.3R & \\
V1151 Sgr & 18:25:23.75 & $-$20:11:59.3 & --   & 2014-04-02 & B600 (1.5 arcsec)
& 4201                   & 17.8R & N \\
V3964 Sgr & 17:49:42.60 & $-$17:23:35.3 & --   & 2014-03-31 & B600 (1.5 arcsec)
& 892                    & 18.0R & N \\
          &             &               &      & 2014-04-22 & B600 (1.5 arcsec)
& 892                    & 18.0R \\
\hline\noalign{\smallskip}
\end{tabular}
\end{minipage}
\end{table*}

For the three targets HS Sge, BS Sgr and V999 Sgr,  $U\!BV\!R$ photometric 
data were taken in 2012 in service mode at the ESO-VLT, Paranal Observatory, 
Chile, using the FOcal 
Reducer/low dispersion Spectrograph \citep[FORS2;][]{appenzelleretal98-3} 
system with the high throughput broad-band filters u\_High, b\_High, v\_High, 
and R\_Special. A series of four frames per filter was obtained. FORS2 
exposures consist of a mosaic of two 2k$\times$4k MIT CCDs that are separated 
by a gap 
with a size of 7 pixels, yielding a total field-of-view of 6.8$\times$6.8 
arcmin$^2$.
The images from the two CCDs are stored in two different files, and we used
the {\sc fsmosaic} routine from the FORS Instrument Mask Simulator ({\sc FIMS})
package to combine them into one image prior to reduction. Using this combined
image instead of reducing each CCD individually introduces a minor statistical 
uncertainty due to the slightly different read-out-noise of the two CCDs 
(2.7 and 3.6 electrons per pixel), 
which, however, is not relevant to our 
results. The images
were corrected with bias and flat frames using the {\sc ccdred} package
of {\sc iraf}. The frames were corrected for the individual telescope offsets
and combined to an average image using a 3$\sigma$ clipping algorithm to 
minimize the effect of bad pixels and 
cosmic rays.
Photometric magnitudes for all 
stars in the fields were extracted using the aperture photometry routines in 
{\sc iraf}'s {\sc daophot} package and the stand-alone {\sc daomatch} and 
{\sc daomaster} routines \citep{stetson92-1}. These magnitudes were calibrated 
using observations of standard stars 
\citep{landolt83-1,landolt92-3,stetson00-2}. The respective calibration
coefficients are gathered in Table \ref{ubvrcalib_tab}. 
Most nights,
only 
one standard field for the $U$ filter was observed, and thus a standard value 
for the extinction was used that was taken from the observatory website%
\footnote{http://www.eso.org/sci/facilities/lasilla/instruments/efosc/inst/zp/.html}.
It is based on measurements taken at La Silla Observatory, about
400 km south of Paranal Observatory. Comparison of the extinction values for
the other filters show that the values should be identical within the
uncertainties.
As we will see in Sections \ref{results_sec} and \ref{v999sgr_sec}, the 
object that was identified as the post-nova in the field of V999 Sgr proved
to be saturated in the photometric data, and thus the photometry of this field 
bears no further relevance for this paper. However, the respective calibration
values are still included in Table \ref{ubvrcalib_tab}, since they serve to
evaluate the consistency of the photometric calibration.

The spectroscopic data were obtained at Gemini-South 
on Cerro Pach\'on, Chile,
in queue mode using the
Gemini Multi-Object Spectrograph \citep[GMOS;][]{hooketal04-1}, at that
time equipped with a CCD array of three 2048x4608 EEV chips. 
Grating B600 with a 1.5 arcsec slit yielded a typical wavelength 
range of 4200 -- 7000 {\AA} and a resolving power of $\sim$800.
A central wavelength at 5600 {\AA} was chosen so that the two 22-pixel-wide 
gaps between the CCDs affect for our purposes comparatively unimportant parts 
of the spectrum. These, however, include the \mbox{He\sc i} lines at 4472
and 5876 {\AA}, respectively.
The reduction of the data was performed using the {\sc gemini -- gmos} 
extension for {\sc iraf} as contained in the {\sc ureka} software 
package\footnote{http://ssb.stsci.edu/ureka/}. This included 
the subtraction of an averaged bias frame and division by a flat-field that 
was previously normalized by fitting a cubic spline of high order to the 
response function. A two-dimensional wavelength calibration solution was
obtained with respect to a CuAr lamp. 
After extraction, the spectra were
corrected for the instrumental function using corresponding data of the
spectrophotometric standard LTT 3218 taken on 2014 March 13. 
This flux calibration can be assumed to be extremely uncertain, because it
relies on a single observation that was performed on a different night than
the observations of the targets, and the latter were not necessarily undertaken
in photometric conditions. In addition to that unknown flux zero-point, we 
found that also the instrumental response function is not well-determined and
shows inconsistencies from one CCD chip to another. This is addressed in
more detail in Section \ref{results_sec}.

We summarize the details on the observations in Table \ref{obslog_tab}, sorted
in order of the variable star name and constellation. The second and
third column of the table contain the coordinates.  For the confirmed novae, 
these were newly determined performing astrometry with Starlink's 
{\sc gaia}\footnote{http://astro.dur.ac.uk/$\sim$pdraper/gaia/gaia.html} 
tool (version 4.4.3) using the US Naval Observatory CCD Astrograph Catalog 
(UCAC) version 3 \citep{zachariasetal10-1} and 4 \citep{zachariasetal13-1}.
The fourth column states the associated root mean square (rms). Coordinates
for the targets that turned out to not be 
post-novae
were taken from 
\citet{downesetal05-1}. Column 5 gives the date of the observations 
corresponding to the start of the night in local time. In column 6 we list
the instrument configuration, i.e.~filter or grating and slit, and column 7
contains the total exposure time. 
Column 8 gives the brightness at the time of the observations with the letter 
at the end identifying the corresponding passband, and the last column presents
an overview on which system was confirmed as a post-nova.
The values for the spectroscopic data were measured
as differential photometry with respect to five comparison stars on the 
acquisition frames which were taken in the GMOS-S $r'$ filter. Calibrated
values were derived by comparison with our previous photometry or with the 
Guide Star Catalog, Version 2.2 \citep{laskeretal08-1}.

\section{Results}
\label{results_sec}

\begin{figure*}
\includegraphics[width=2.0\columnwidth]{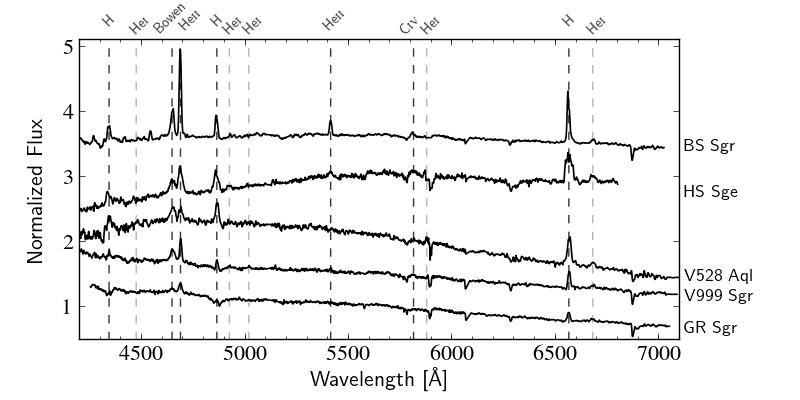}
\caption[]{Spectra of the confirmed post-novae. 
Vertical lines mark the principal emission lines as labelled.
}
\label{novasp_fig}
\end{figure*}

\begin{figure*}
\includegraphics[width=2.0\columnwidth]{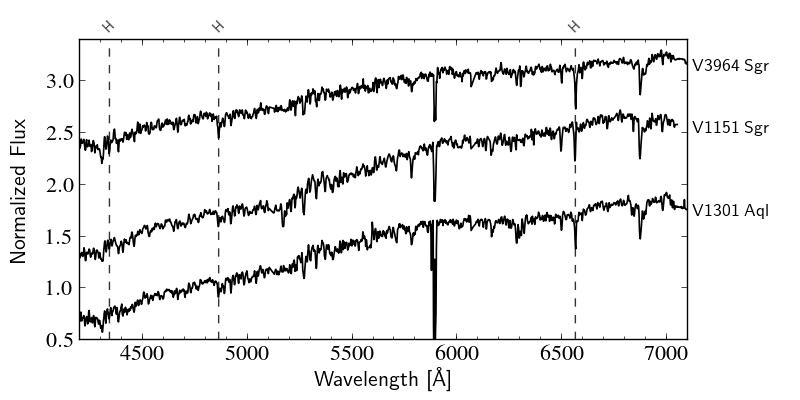}
\caption[]{Spectra of the three targets that turned out not to be post-novae.
Vertical lines mark the positions of hydrogen lines.
}
\label{nonovasp_fig}
\end{figure*}

\begin{table*}
\begin{minipage}{16cm}
\caption[]{Equivalent widths in angstroms of the principal emission lines.
Values in square brackets refer to components within absorption troughs.
}
\label{eqw_tab}
\begin{tabular}{@{}llllllllllll}
\hline\noalign{\smallskip}
Object    &  \multicolumn{3}{c}{Balmer} 
& \multicolumn{4}{c}{He\,{\sc i}} & Bowen & \multicolumn{2}{c}{He\,{\sc ii}} 
& C\,{\sc iv}\\
          & 4340   & 4861       & 6563        & 4922   & 5016   
& 5876    & 6678   & 4645 & 4686 & 5412   & 5812 \\
\hline\noalign{\smallskip}
V528 Aql  & 1.8(3) & 4.0(4) & 13.3(7)     & --     & --     
& --$^1$  & 2.5(2) & 4.6(5) & 2.9(6)  & --     & --     \\
HS Sge    & 6.0(6) & 7.6(3) & 14.6(3) & 1.4(2) & 0.3(1)
& --$^1$  & 2.5(2) & 17.2(5)$^2$ & 17.2(5)$^2$ & 1.5(2) & --     \\
BS Sgr    & 3.5(2) & 4.7(1)$^3$ & 12.2(1)$^3$ & 0.4(1) & 0.2(1) 
& --$^1$  & 1.4(1) & 7.1(1) & 15.1(2) & 2.8(1) & 1.2(3) \\
GR Sgr    & [0.4(1)] & [1.0(2)] & 2.5(1)      & --    & --       
& --      & 0.8(1) & 0.6(1) & 1.8(3)  & --     & --      \\
V999 Sgr  & [0.6(1)] & [1.7(1)] & 4.7(2)      & 0.2(1) & 0.3(1)        
& --$^1$  & 1.1(1) & 2.7(3) & 3.1(1)  & 0.5(1) & 0.8(2)  \\
\hline\noalign{\smallskip}
\end{tabular}
\\
$^1$ Present, but too distorted for measurement.\\
$^2$ Bowen/\mbox{He\sc ii} blend.\\
$^3$ Blended with \mbox{He\sc ii} emission.
\end{minipage}
\end{table*}

\begin{figure}
\includegraphics[width=\columnwidth]{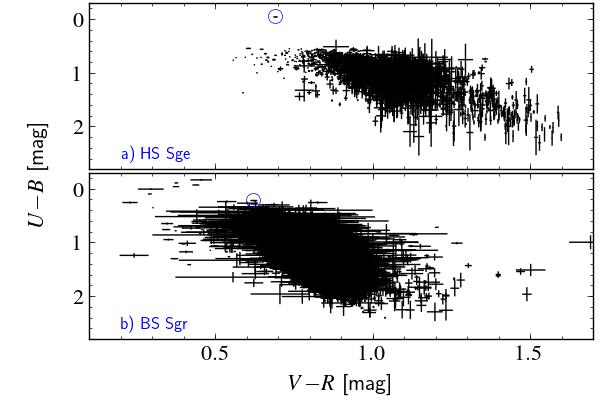}
\caption[]{
Colour-colour diagrams for the two post-nova fields with $U\!BV\!R$ 
data.
The confirmed post-novae are marked with a circle.}
\label{ubvr_fig}
\end{figure}

%The spectra of the five confirmed post-novae are presented in 
%Fig.~\ref{novasp_fig}, those of the three misidentifications in 
%Fig.~\ref{nonovasp_fig}. 

While in the subsequent chapters we present the results on the individual 
targets, in the following we briefly describe our analysis methods that 
apply to all objects.

We analyse the spectra of the 
confirmed post-novae with respect to the spectral energy distribution (SED)
and the properties of the principal emission lines. To measure the equivalent 
widths $W_\lambda$, three wavelength ranges were specified manually. They 
consisted of the central range that fully contains that line, and of two small
adjacent sections to the blue and the red that served to compute a linear fit 
to the continuum. For convenience, we use positive values for $W_\lambda$. 
This is unambiguous since we exclusively refer to emission lines. The 
corresponding errors were estimated with a Monte Carlo simulation by adding 
random noise to the data and repeating the 
equivalent width measurement a thousand times.

In order to recover the intrinsic SED, we 
employ the values for interstellar 
extinction $E(B\!-\!V)$ from 
\citet{schlafly+finkbeiner11-1}, collected on NASA's Infrared Science Archive 
(IRSA) web pages, to deredden the spectra. We use the relations 
derived by \citet*{cardellietal89-1} as implemented in {\sc iraf}, and
a standard value for the ratio of the total to the selective extinction 
$R(V) = A(V)/E(B\!-\!V) = 3.1$. The corresponding values are available as the 
average extinction within a 2$\times$2 deg$^2$ field. This represents a severe 
limitation as it does not take into account potential small-scale variations 
in the dust structure, nor the actual amount of absorbing material in the line 
of sight to the target. Since CVs are intrinsically rather faint 
\citep[typically, $M_V > 4$ mag; e.g.][]{patterson84-1}, in general
the correction for the interstellar extinction will represent an upper limit.
An additional uncertainty regarding the intrinsic SED 
is introduced by obtaining the flux calibration on the basis of a single 
standard star that was observed on a different night than the targets. 
Furthermore, the
observations were not conducted at a parallactic angle and
thus might be affected by chromatic slit loss due to differential refraction,
since Gemini-South is not equipped with an atmospheric dispersion corrector.
As a measure for the SED, we derive the exponent
$\alpha$ of a power law $F \propto \lambda^{-\alpha}$ that was fitted to
the spectra, restricting the continuum to wavelengths
5000--7000 {\AA} and masking strong emission and absorption lines.
During that analysis we found that most of the post-novae required a 
two-component fit while previously this represented an exception 
\citepalias{tappertetal12-1,tappertetal14-1}. Furthermore, in every case the
two slopes differed in the same direction, with the red part of the spectrum 
requiring a steeper slope than the blue part, and the points that 
separated the two slopes were found to be within 5850--5900 {\AA} which marks
the region that is affected by the gap between two CCDs (see Section
\ref{obs_sec}). All this suggests a systematic effect in the definition of the 
instrumental response function, and potentially a difference in the spectral 
efficiency of these two CCDs. We have thus further divided the 
fitting range into a blue (5000 {\AA} $< \lambda <$ 5870 {\AA}) and a red 
(5870 {\AA} $< \lambda <$ 7000 {\AA}) part that are fitted individually. The 
consequences for the interpretation of the SEDs are further discussed
in Section \ref{disc_sec}.

Finally, to estimate the spectral type of the targets that turned out not
to be post-novae, as well as for a few other objects, we used the standard
star libraries of \citet*{jacobyetal84-1}, \citet{pickles85-2} and
\citet{silva+cornell92-1}.

%Finally, for the two novae with $U\!BV\!R$ photometry Fig.~\ref{ubvr_fig} 
%presents the respective colour-colour diagrams, and Table \ref{ubvr_tab} 
%details the corresponding values.

\subsection{V528 Aquilae = Nova Aql 1945}
\label{v528aql_sec}

As reported by \citet{duerbeck87-1}, this fast nova was discovered on 1945,
August 26 by Bertaud and then independently two nights later by Tamm. 
\citet*{stropeetal10-1} analysed the decline curve and classify it as
`smooth', although it appears to be modulated with small-amplitude 
oscillations \citep*{bianchinietal92-1}. 

Our observations show that the post-nova is part of a 
close visual binary. Astrometry gives the position of the 
neighbour
to
RA = 19$^\mathrm{h}$19$^\mathrm{m}$19$\fs$12, 
Dec.~= +00$\degr$37$\arcmin$53$\farcs$8, and thus approximately 0.8 arcsec
north-east of the post-nova. The 
neighbour
is also the slightly brighter
object. Comparison with the GSC 2.2 catalogue yields $R = 18.56(31)$ for this
star, and $R = 18.66(31)$ mag for the post-nova. Note that above errors are 
dominated by the uncertainty in the zero-point, while the precision for
the differential photometry is $\sim$0.01 mag. Our comparison of the
spectrum of the 
close neighbour
with standard star libraries indicates a spectral
class of late G to early K. Previous photometry of V528 Aql, e.g.~ by 
\citet{szkody94-2}, is likely to have included both components, thus yielding 
a larger brightness and redder colours.
Based on the present photometric data $m_\mathrm{post-nova} = 18.7$ mag and
the maximum brightness $m_\mathrm{max} = 6.9$ mag from \citet{stropeetal10-1}
we calculate the eruption amplitude 
$\Delta m = m_\mathrm{post-nova} - m_\mathrm{max} = 11.8$ mag. 

Due to the close apparent proximity, the visual binary is not resolved in our
spectroscopic data. This made it necessary to extract the spectra
simply as the average of 10 CCD rows containing a part of the point-spread 
function that was selected to minimize contamination by the neighbour,
without the usually performed tracing and fitting the spatial position as a 
function of wavelength.
However, this turned out not to be a severe limitation since the 
variation in the position resulted to $\sim$1 pixel over the whole wavelength 
range. 
The extracted
spectrum of the post-nova is presented in 
Fig.~\ref{novasp_fig}. It shows moderately strong emission of the Balmer
series as well as \mbox{He\sc i} (Table \ref{eqw_tab}). 
The Bowen blend and \mbox{He\sc ii} are
also present, but not particularly strong. No other \mbox{He\sc ii} or
carbon lines can be identified. 
The difference in the slope parameters $\alpha$ is particularly
pronounced in this object, amounting to $\sim110~\sigma$ (Table 
\ref{prop_tab}). 
We suspect that this is caused by a certain level of contamination by the 
close neighbour. 

Overall, the spectrum 
indicates a non-magnetic CV
with a, for post-novae, medium high mass-transfer rate. The lines
are neither particularly broad nor narrow, pointing to a medium system
inclination. In our individual spectra a displacement of the emission lines
especially of the first spectrum with respect to the other two is clearly
detected. This is thus a suitable target for spectroscopic time series
observations to derive the orbital period, with the caveat that such data
have to be taken in very good seeing conditions in order to properly separate
the two components of the visual binary.

\subsection{V1301 Aquilae = Nova Aql 1975}
\label{v1301aql_sec}

\begin{table}
\caption[]{Coordinates (J2000.0) and magnitudes of the possible post-nova 
candidates for V1301 Aql.}
\label{v1301aql_tab}
\begin{tabular}{@{}lllll}
\hline\noalign{\smallskip}
RA          & Dec.      & $R$ (mag)   & Remarks \\
\hline\noalign{\smallskip}
19:17:55.24 & +04:47:18.4 & --        & \citet{duerbeck87-1} \\
19:17:55.39 & +04:47:13.4 & --        & \citet{wild75-2}\\
19:17:55.31 & +04:47:17.4 & 18.91(36) & `D' \\
19:17:55.29 & +04:47:13.3 & 20.02(36) & `W1' \\
19:17:55.34 & +04:47:12.9 & 19.98(36) & `W2' \\
19:17:55.37 & +04:47:15.4 & 22.18(41) & `F' \\
\hline\noalign{\smallskip}
\end{tabular}
\end{table}

This is a comparatively recent nova that was discovered by \citet{wild75-2}.
The date of maximum brightness as well as the eruption magnitude were
later corrected by \citet{howarth76-3} to 1975 June 4 and 10.3 mag, 
respectively. The authors also present the decline light curve. 
\citet{duerbeck87-1} classifies the object as a fast nova with
a time of decline by 3 mag $t_3 = 35$ d. Infrared data and a red optical 
spectrum 45 d after the discovery are included in \citet{vrbaetal77-1}, and
\citet*{peschetal75-1} describe the spectrum 95 d after eruption.

We have observed the object marked in \citet{duerbeck87-1} and
\citet{downesetal05-1}. However, the spectrum does not resemble that of a
CV (Fig.~\ref{nonovasp_fig}). Instead, after dereddening we find very good
agreement with the data of HD 29050 that \citet{jacobyetal84-1} classify as
a G9V--K1V star. Photometry on our acquisition frame gives $R = 18.91(36)$ mag,
and thus the target also appears as too bright, because the implied eruption
amplitude of $\sim$8.6 mag would be unusually small for a fast nova
\citep{warner95-1},
although it is well within the range of recurrent novae \citep{schaefer10-2}.

The published eruption spectra 
indicate
that V1301 Aql was a nova, and since the
decline light curve has been comparatively well covered, we do not expect a 
very large
uncertainty with respect to its position. Comparing the coordinates given in 
\citet{duerbeck87-1} and \citet{downesetal05-1} to those of the original 
discovery by \citet{wild75-2} yields a discrepancy of about 6 arcsec. On our 
acquisition frame we find a visual binary close to the 
\citet{wild75-2} coordinates, its components marked with `W1' and `W2' on our 
finding chart in Fig.~\ref{fcs1_fig}, where the original candidate is marked as 
`D'. Both components are also about 1 mag fainter than `D' and would thus
agree somewhat better with the classification of a fast nova. Another possible
candidate is a very faint object situated between `D' and `W2', marked
as `F'. Its brightness would fit best that expected of a fast nova.
The coordinates and derived magnitudes are summarized in Table 
\ref{v1301aql_tab}. Further photometric and spectroscopic data will be
necessary to recover the nova, but for now the above three objects appear
as the best candidates.

\subsection{HS Sagittae = Nova Sge 1977}
\label{hssge_sec}

\begin{table}
\caption[]{Results of the $U\!BV\!R$ photometry.}
\label{ubvr_tab}
\begin{tabular}{@{}lllll}
\hline\noalign{\smallskip}
Object   & $V$       & $U\!-\!B$   & $B\!-\!V$ & $V\!-\!R$ \\
\hline\noalign{\smallskip}
HS Sge   & 19.99(03) & $-$0.06(01) & 0.88(01)  & 0.69(01) \\
BS Sgr   & 17.90(05) &    0.21(01) & 0.69(01)  & 0.62(01) \\
\hline\noalign{\smallskip}
\end{tabular}
\end{table}

\begin{figure}
\includegraphics[width=\columnwidth]{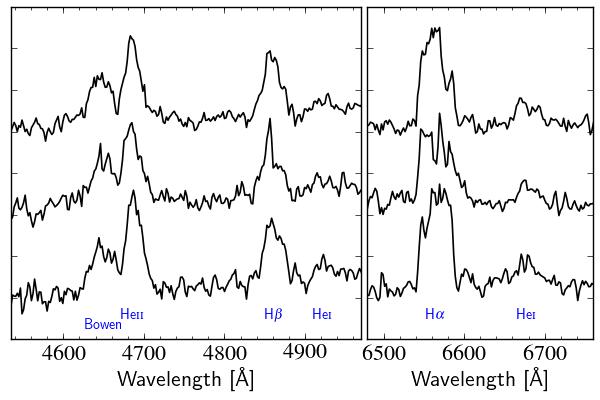}
\caption[]{Close-up on the individual spectra of HS Sge (first to last in
time from bottom to top). 
The time difference between subsequent spectra is 31 min.
Emission lines are labelled at the bottom of the figure.
}
\label{hssgelprof_fig}
\end{figure}

Another recent nova, HS Sge was discovered 1977, January 7 at a visual
brightness of 7.2 mag \citep*{milbournetal77-1}. The decline light curve was 
analysed by \citet{stropeetal10-1} who classify it as type `P' (plateau) and
derive a fast decline time $t_3 = 21$ d. \citet{szkody94-2} present photometric
data of the post-nova and finds $V = 19.4$ mag. However, the finding chart
in \citet{downesetal05-1} marks a blend of several stars so that the precise
position of the post-nova was not sufficiently well known to obtain a spectrum.

In the $U\!BV\!R$ colour-colour diagram one object close to the reported 
coordinates stands out in being positioned far away from the bulk of the field 
stars (Fig.~\ref{ubvr_fig}, Table \ref{ubvr_tab}). 
The spectrum confirms the nova, presenting
moderately strong Balmer emission lines, \mbox{He\sc i}, the Bowen blend,
and a strong contribution of \mbox{He\sc ii}, which can be detected both at
$\lambda$4686 and $\lambda$5412. The profile of the H$\alpha$ emission
line is especially interesting in that it contains unusually steep wings and
appears to include several peaks. In Fig.~\ref{hssgelprof_fig} we show a
close-up of several lines in the three individual spectra. The presence of
an additional narrow emission component that moves within the broader line
profile can be clearly identified in the Balmer lines, most prominently in the 
second spectrum. At least the blue wing of the H$\alpha$ line still appears
unusually steep. One might thus suspect that this is caused by the presence
of another emission component. However, other than in BS Sgr (Section
\ref{bssgr_sec}), the \mbox{He\sc ii} emission lines corresponding to the
transition to the third excitation level are weak, and only the line at
$\lambda$5412 can be identified with certainty. This excludes a significant
contribution of the $\lambda$6560 line to the H$\alpha$ profile. Furthermore, 
like the other emission lines, H$\alpha$ is found to be centred rather 
symmetrically on its rest wavelength, which makes a distortion due to other
lines rather unlikely. Since a comparatively short time has passed since the
nova eruption, one could suspect that the line profile is due to additional
emission from the ejected matter. However, on the 2D spectroscopic image we do 
not find any evidence for the presence of a shell. Clearly, more data are 
needed to investigate the line profile in more detail. At the very least, the 
broad and variable line profiles
point to a high system inclination, which makes HS Sge an attractive target 
for time series observations.

\citet{stropeetal10-1} suspect that 
novae
with plateau-type decline light curves
are either recurrent novae 
\citep[see also][]{pagnotta+schaefer14-1}
or belong to the V1500 Cyg class, i.e.~contain
a highly magnetized white dwarf. The large eruption amplitude $\Delta m =
12.8$ mag and the strong presence of \mbox{He\sc ii} lets the latter scenario
appear as the more likely one. However, the shape and width of especially the
\mbox{He\sc i} $\lambda$6678 emission line indicates the existence of at least
an outer disc, 
so that the magnetic field of the white dwarf can be assumed
to be weaker than in polars.

\subsection{BS Sagitarii = Nova Sgr 1917}
\label{bssgr_sec}

\begin{figure}
\includegraphics[width=\columnwidth]{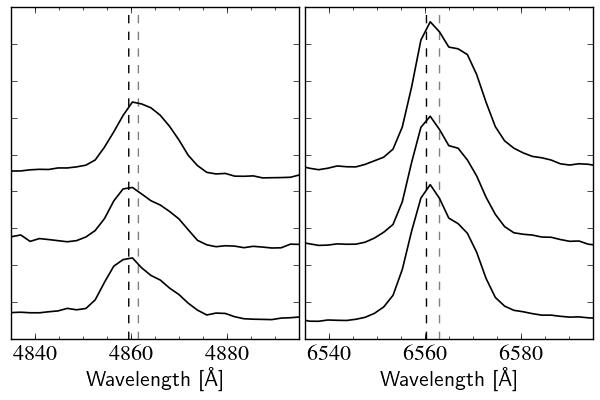}
\caption[]{Close-up on the individual spectra of BS Sgr (first to last in
time from bottom to top). The dashed lines mark the rest wavelengths of
the \mbox{He\sc ii} (black) and the Balmer (grey) emission.
The time difference between subsequent spectra is 31 min.
}
\label{bssgrlprof_fig}
\end{figure}

The nova was discovered by \citet{innes17-2} on a plate from 1916 July 17
at a magnitude of 11.8, reaching 10.5 mag two nights later. An independent
and more comprehensive study by \citet{cannon23-5} showed that the object
rose slowly to a maximum brightness 
of
9.2 mag exactly one year after
its discovery, followed by a  slow decline with $t_3 \sim 700$ d 
\citep{duerbeck87-1}. Perhaps due to the scarceness of the data, BS Sgr is not 
part of the long-term light-curve analysis of \citet{stropeetal10-1}, but the 
observed behaviour suggests that it belongs to the small group of novae with 
`flat-topped' (F class) light curves. The authors find four such systems
within their sample of 93 novae. The physical reason behind the prolonged
plateau near maximum brightness is unknown. \citet{duerbeck+seitter87-1}
find $V = 15.45$, $B\!-\!V = 1.10$ and $V\!-\!R = 0.70$ from spectrophotometric 
data on the post-nova taken between 1983 and 1985, and describe the spectral
range between 3900--7000 {\AA} as featureless apart from the
\mbox{He\sc ii} and Bowen blend, and perhaps \mbox{C\sc ii} $\lambda$4267.
However, as we will see below in more detail, neither the brightness and 
colours, nor the description of the spectrum agrees well with the object that
we identify as the likely post-nova (Table \ref{ubvr_tab}, 
Fig.~\ref{novasp_fig}). The latter is located about 5 arcsec south-east of the
object indicated in the finding chart by \citet{duerbeck87-1}. It appears
possible that due to the closeness of the objects, the spectrum described by 
\citet{duerbeck+seitter87-1} is a result of the slit being centred on the 
object marked in \citet{duerbeck87-1}, but, depending on the orientation of 
the slit and the seeing conditions, includes some contribution by the
post-nova, thus explaining the apparent presence of the blue emission lines.
Subsequent studies, e.g.~by \citet*{ringwaldetal96-3} (who remark that the
colours are that of a reddened K star), \citet{hoardetal02-1} and 
\citet{saitoetal13-2}, can be assumed to refer to the object identified in
\citet{duerbeck87-1}.

The colour-colour diagram, while containing several blue objects, does not
immediately indicate the nova (Fig.~\ref{ubvr_fig}). The broad distribution
of colours suggests that the likely reason for this is a certain patchiness
of the interstellar extinction in this region. However, 
we found the bluest object within the central 1 arcmin of the field to be
located only 5 arcsec off the reported coordinates of the nova eruption, which 
made it a promising target for the spectroscopic follow-up observations. 

The spectrum 
confirms the CV. It
is remarkable in that it presents an unusually strong contribution
from \mbox{He\sc ii}. First, the $\lambda$4686 line, which originates from the 
transition to the second excitement level, represents by far the strongest 
emission line in BS Sgr. Secondly, also all lines that originate from the 
transition to the third level are detected, at $\lambda\lambda$4542, 4859,
5412 and 6560 {\AA} (Fig.~\ref{novasp_fig}). The two lines that are close to 
the Balmer lines are not resolved, but visibly distort the line profile. In
Fig.~\ref{bssgrlprof_fig} we show a close up on the two Balmer -- 
\mbox{He\sc ii} blends in the three individual spectra. We notice that from
the first to the third spectrum the Balmer components become stronger
indicating that the region where that emission originates moves further into
the line of sight. Potentially present different radial velocities of the
two components could further contribute to the variation of the line profiles.

Prominent \mbox{He\sc ii} emission usually indicates the presence of a highly
magnetic white dwarf primary. We can probably exclude the possibility that
BS Sgr is a polar, because the absence of an accretion disc should be
reflected in a large 
eruption amplitude 
\citep[e.g.~the best candidate for such a system, V1500 Cyg, 
has $\Delta m = 16$ mag;][]{stropeetal10-1}, and with $\Delta m =
8.7$ mag BS Sgr is clearly a nova with a rather small eruption amplitude.
It is thus more likely to be an intermediate polar. On the other hand, the
emission lines are comparatively narrow (Table \ref{prop_tab}), while at the
same time the variation in the line profiles mentioned above -- if due to
an orbital effect -- suggests that the system is seen at 
a medium to high
inclination. This makes it rather unlikely that the emission lines originate
in an accretion disc.
A likely alternative is that they are located on the potentially irradiated
side of the secondary star \citepalias[see e.g.][where we find such an
emission component in the old nova V728 Sco]{tappertetal13-1}.

\subsection{GR Sagitarii = Nova Sgr 1924}
\label{grsgr_sec}

\begin{figure}
\includegraphics[width=\columnwidth]{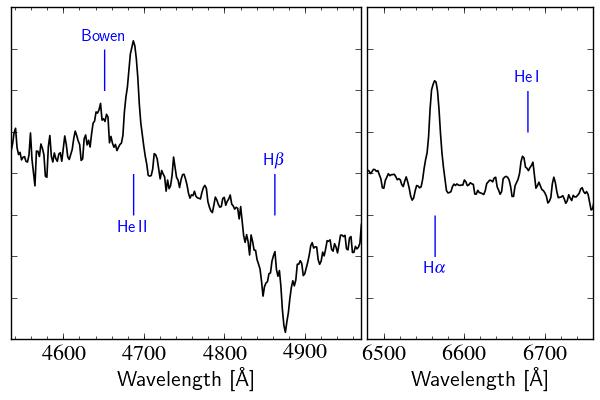}
\caption[]{Close-up on the line profiles in the spectrum of GR Sgr.}
\label{grsgrlprof_fig}
\end{figure}

\citet{woodsie27-3} discovered the nova on a plate from 1924 April 30 at 11.4 
mag. Based on subsequent data she suspected that maximum brightness had been 
missed. \citet{duerbeck87-1} agrees that the nova had 
erupted several 
months before its discovery and estimates that the object at maximum
was about 4 mag brighter. The quiescent brightness is also reported by Woods
as $\sim$16.6 mag both before (1923 September) and after (1925 October) the
eruption. Further, near-infrared, photometric data are given by 
\citet{hoardetal02-1} and \citet{saitoetal13-2}. Although the post-nova is
comparatively bright, the published spectrum by \citet{ringwaldetal96-3} shows 
little detail, apart from a few, weak, emission lines, leading the authors to
only tentatively confirm the object as a nova. The previous 
description of a spectrum by \citet{duerbeck+seitter87-1} is equally 
unsatisfactory.

Our own spectroscopic data in Fig.~\ref{novasp_fig} show a blue
continuum with the higher Balmer lines being embedded in the absorption troughs
of an optically thick disc. There is even a hint of the presence of absorption
in H$\alpha$ (Fig.~\ref{grsgrlprof_fig}). In contrast, the Bowen blend, 
\mbox{He\sc ii} and the only detectable \mbox{He\sc i} line at $\lambda$6678,
are purely in emission. 

Photometry of 
the acquisition frame yields $R = 15.7(8)$ mag, which within the uncertainties 
corresponds to the same brightness as the spectrophotometrically derived values 
by \citet{duerbeck+seitter87-1} and \citet{ringwaldetal96-3}. Clearly, GR Sgr 
still remains in a state of a very high accretion rate, 90 yr after its 
eruption.

\subsection{V999 Sagitarii = Nova Sgr 1910}
\label{v999sgr_sec}

\citet{pickering10-55} reports the discovery of this object by W.~Fleming
on photographic plates from 1910 March 21 to June 10, where it reached a 
maximum brightness of 7.8 mag. The further photometric behaviour is described 
e.g.~by \citet{swope40-4}. \citet{duerbeck87-1} classifies it as a
slow nova with $t_3 = 160$ d. \citet{duerbeck+seitter87-1} describe the
post-nova spectrum as a blue continuum with weak emission lines. They
derive spectrophotometric magnitudes and colours to $V = 16.60$ mag,
$B\!-\!V = 0.75$, $V\!-\!R = 0.35$. However, in strong contrast to that 
description, a spectrum presented by \citet{ringwaldetal96-3} showed a 
featureless red continuum, with corresponding spectrophotometric values 
$V = 19.1$ mag, $B\!-\!V = 3.4$, $V\!-\!R = 1.2$. \citet{saitoetal13-2}
do not detect the post-nova in the near-infrared.

Since the identification of the post-nova thus seemed unclear, we took
$U\!BV\!R$ photometry of the field. The respective colour-colour diagram did
not present any interesting star near the reported position. However, a visual
inspection of the images showed that the object in the centre of the knot
of stars marked by \citet{downesetal05-1}, just above the saturation limit of 
our observations, appeared blue, and was thus selected for follow-up 
spectroscopy. 

The spectrum in Fig.~\ref{novasp_fig} confirms the nova. Its appearance is
somewhat similar to that of GR Sgr (Section \ref{grsgr_sec}) in that the
higher Balmer lines are in absorption troughs, indicating a high mass-transfer
rate. In contrast to GR Sgr, the troughs are not as profound, the blue
lines of the \mbox{He\sc i} series are clearly detected, and the higher
excitation lines of \mbox{He\sc ii} $\lambda$5412 and \mbox{C\sc iv}
$\lambda$5812 are present. The lines are narrow (Table \ref{prop_tab}), so 
that the system inclination can be suspected to be comparatively low. In 
agreement with this, an examination of the three individual spectra does not 
show any evidence for radial velocity variations.

The IRSA web page lists a very high extinction value for the region around
V999 Sgr of $E(B\!-\!V) = 1.70(11)$. Correcting for this value permits to
fit the SED with an exponent $\alpha = 6.5$. However, considering the 
spectral appearance, this value appears exaggerated 
\citepalias[see also ][]{tappertetal12-1,tappertetal14-1}, and thus applying
the average extinction here very likely results in an overcorrection of the
reddening. 

Finally, the results from \citet{ringwaldetal96-3} and \citet{saitoetal13-2}
raise the possibility that V999 Sgr shows large variability. However,
the post-nova is located in a comparatively crowded field, and the
previously published finding charts left ample room for ambiguity. Thus,
\citet{ringwaldetal96-3} likely observed a different object. Similarly, the
coordinates used by \citet{saitoetal13-2} are roughly 1.6 arcsec off the
actual position, and thus the post-nova is just outside their identification
limit of 1 arcsec, with the large number of stars in the vicinity additionally 
preventing an unambiguous choice. 
A closer inspection of the publicly
available data from the Vista Variables in the V\'ia L\'actea (VVV) survey
\citep{saitoetal12-1} shows that the post-nova is indeed visible on all
$ZY\!H\!J\!K$ images, and that it corresponds to the VVV source 515510491978. 
For this object, $J$ and $K$ data tabulated magnitudes are available, with 
$J = 16.06(11)$ mag and $K = 16.19(31)$ mag for an aperture radius of 1.0 
arcsec.

\subsection{V1151 Sagitarii = Nova Sgr 1947}
\label{v1151sgr_sec}

\begin{table}
\caption[]{Coordinates (J2000.0) and magnitudes of the possible post-nova 
candidates for V1151 Sgr.}
\label{v1151sgr_tab}
\begin{tabular}{@{}lllll}
\hline\noalign{\smallskip}
RA          & Dec.      & $R$ (mag)   & Remarks \\
\hline\noalign{\smallskip}
18:25:23.79 & $-$20:11:59.3 & --        & \citet{duerbeck87-1} \\
18:25:20    & $-$20:11:51   & --        & \citet{mayall49-1}\\
18:25:23.86 & $-$20:11:58.3 & 17.80(61) & `D' \\
18:25:23.79 & $-$20:11:59.4 & 18.87(61) & `1' \\
\hline\noalign{\smallskip}
\end{tabular}
\end{table}

According to \citet{duerbeck87-1}, this little-studied nova reached a maximum 
photographic brightness of 11.1 mag on 1947 August 11, while the original
discovery note by \citet{mayall49-1} gives $m_\mathrm{max} = 10$ mag, without
specifying the date. Based on an unpublished light curve by J.~Warren, 
Duerbeck classifies it as a slow nova with $t_3 = 135$ d. To our knowledge,
there are no further observations of this nova other than the data
from the 2 Micron All-Sky Survey provided by \citet{hoardetal02-1}.

We took a spectrum of the object marked and described (`north object of close
pair') by \citet{downesetal05-1}, but found it to show a red continuum
without any emission lines (Fig.~\ref{nonovasp_fig}). 
A comparison of this spectrum, corrected for interstellar extinction
($E(B\!-\!V) = 0.81$), with the standard star catalogue of 
\citet{jacobyetal84-1} showed it to be in good agreement with
a spectral type $\sim$K3V. 
However, the acquisition frame shows a $\sim$1 mag fainter object 
close to our target, whose coordinates agree even better with those given by 
\citet{duerbeck87-1}, and which thus appears as a likely candidate for the
post-nova. We have marked this object as `1' and the original candidate as 
`D' on the finding chart in Fig.~\ref{fcs2_fig}, and summarize the data in 
Table \ref{v1151sgr_tab}. As a caveat, we remark that the revised coordinates
from \citet{duerbeck87-1} disagree considerably (by almost 1 arcmin) with 
those originally reported by \citet{mayall49-1}. It might thus be advisable
to first confirm photometrically that `1' is a good candidate before
`risking' a spectrum.

\subsection{V3964 Sagitarii = Nova Sgr 1975}
\label{v3964sgr_sec}

\begin{table}
\caption[]{Coordinates (J2000.0) and magnitudes of the possible post-nova 
candidates for V3964 Sgr.}
\label{v3964sgr_tab}
\begin{tabular}{@{}lllll}
\hline\noalign{\smallskip}
RA          & Dec.      & $R$ (mag)   & Remarks \\
\hline\noalign{\smallskip}
17:49:42.39 & -17:23:35.7 & --        & \citet{duerbeck87-1} \\
17:49:42.0  & -17:23:21   & --        & \citet{lundstrom+stenholm77-1}\\
17:49:42.62 & -17:23:36.1 & 17.95(49) & `D' \\
17:49:42.42 & -17:23:34.8 & 19.99(49) & `1' \\
\hline\noalign{\smallskip}
\end{tabular}
\end{table}

The nova was discovered on an objective prism plate from 1975 June 8, by 
\citet{lundstrom+stenholm76-1,lundstrom+stenholm77-1}. The authors estimate
the brightness to $\sim$8.5 mag. Based on the spectral appearance they
suggest that the real maximum was missed and that the nova had already
declined by $\sim$3 mag. A search on Sonneberg plates by \citet{huth76-1}
revealed an earlier detection on 1975 June 1, at a brightness of 9.4 mag.
If the suspicion by \citet{lundstrom+stenholm77-1} is correct, this would
make it a pre-maximum detection. The fact that the nova was not detected to
a plate limit of 11 mag already on 1975 July 5, in any case classifies it as
a fast nova. \citet{duerbeck87-1} estimates $t_3 = 32$ d. 
\citet{pattersonetal81-2} include the object in a list of potential
WZ Sge-type dwarf novae, however, without providing a more detailed
justification. 
The near-infrared data listed in \citet{hoardetal02-1}
suggest $R > 18$ mag \citep[using typical colour values from][]{szkody94-2} and
thus -- as we will see below -- likely correspond to the wrong star.

The object marked by \citet{downesetal05-1} turned out to be a knot of
several stars. Our spectroscopic observations were centred on the brightest
of those, marked as `D' in the finding chart (Fig.~\ref{fcs2_fig}). This,
however, is not the post-nova system, but a red main-sequence star. From
comparison of the dereddened ($E(B\!-\!V) = 0.56$) data with 
\citet{jacobyetal84-1}
and \citet{pickles85-2} we estimate a spectral type $\sim$K1V. The object was
observed twice, because during the first time the slit was not well centred
on the target. A visual inspection of the two-dimensional spectroscopic data 
showed that during that first observation weak H$\alpha$ emission could be
detected north-west of the target. This emission was not spatially 
resolved, leaving three faint objects as potential sources. However, during
the second observation this emission was not detected. Comparison of the
respective slit images finally revealed one object that was found to be 
partially in the slit in the first data set, and out of the slit in the 
second. 
This object, which is thus very likely to be the post-nova, is marked
as `1' in our finding chart (Fig.~\ref{fcs2_fig}).

\section{Discussion}
\label{disc_sec}

\begin{table*}
\begin{minipage}{2.0\columnwidth}
\caption[]{Several properties of the five post-novae. See Section 
\ref{disc_sec} for details.}
\label{prop_tab}
\begin{tabular}{@{}lllllllllll}
\hline\noalign{\smallskip}
Object & $m_\mathrm{max}$$^1$ & $m_\mathrm{min}$ & $\Delta m$ & $\Delta t$ 
& $t_3$ & $E(B\!-\!V)$ & $\alpha$$^2$ & $W_\lambda(\mathrm{H}\alpha)$ &
\multicolumn{2}{c}{FWHM({H}$\alpha$)} \\
 & (mag) & (mag) & (mag) & (yr) & (d) & (mag) & & ({\AA}) & ({\AA}) 
& (km s$^{-1}$) \\
\hline\noalign{\smallskip}
V528 Aql & 6.9$V$ & 18.7 & 11.7 & 69  & 37  & 0.39 & 2.96(03)/5.15(04)
& 13 & 20 & 915 \\
HS Sge   & 7.2$v$ & 20.0 & 12.8 & 35  & 21  & 1.20 & 2.92(03)/3.31(01) 
& 15 & 35 & 1600 \\
BS Sgr   & 9.2$p$ & 17.9 &  8.7 & 95  & 700 & 0.33 & 1.19(02)/1.81(02) 
& 12$^3$ & 14$^3$ & 640$^3$ \\
GR Sgr   & $<$11.4$p$ & 15.7 & $>$5.1 & 90 & -- & 0.46 & 2.36(02)/2.97(02)
& 3 & 15 & 685 \\
V999 Sgr & 7.8$p$ & 16.3 &  8.5 & 104 & 160 & 1.70 & 6.69(02)/6.61(02)
&  5 & 13 & 595 \\
\hline\noalign{\smallskip}
\end{tabular}
\\
$^1$ $p$: photographic, $V$: $V$-band, $b$: blue, $v$: visual.\\
$^2$ First value for 5000 {\AA} $< \lambda <$ 5870 {\AA}, second for 
5870 {\AA} $< \lambda <$ 7000 {\AA} (see text).\\
$^3$ Blend.
\end{minipage}
\end{table*}

\begin{figure}
\includegraphics[width=\columnwidth]{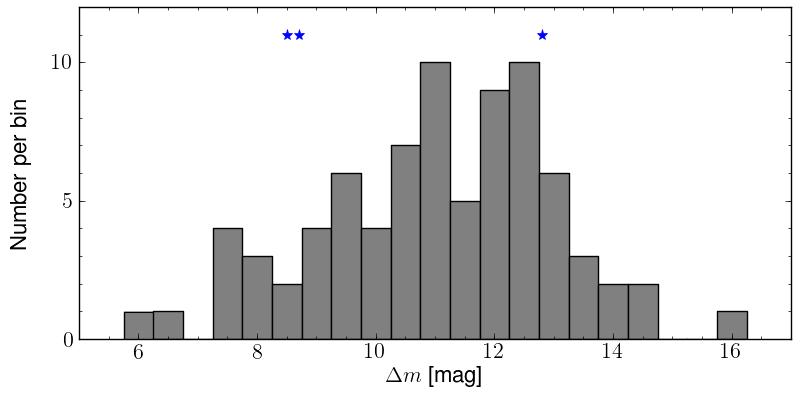}
\caption[]{Distribution of the eruption amplitudes $\Delta m$ of novae that 
erupted before 1986. HS Sge, BS Sgr and V999 Sgr are not included in the
histogram. Their amplitudes are indicated at the top of the distribution.}
\label{amphis_fig}
\end{figure}

The confirmed post-novae of this work, 
V528 Aql, HS Sge, BS Sgr, GR Sgr and 
V999 Sgr, cover a broad range of parameters, which we have summarized in 
Table \ref{prop_tab}: the apparent
maximum brightness $m_\mathrm{max}$, the brightness $m_\mathrm{min}$ during
our observations, the resulting 
eruption amplitude $\Delta m = 
m_\mathrm{min} - m_\mathrm{max}$, with $m_\mathrm{min} \equiv 
m_\mathrm{post-nova}$ (as already used in Section \ref{v528aql_sec}),
the `age' of the post-nova $\Delta t$, 
i.e.~the time that has passed between maximum brightness and our observations,
the time
of decline to three magnitudes below maximum $t_3$, the interstellar
extinction $E(B\!-\!V)$ that was used for the dereddening, 
the exponents 
$\alpha$ of the power laws that were fitted to the dereddened `blue' and
`red' SED with the errors representing the standard deviation of the slope of 
the fit 
and finally 
the equivalent width $W_\lambda$ and the full width at half-maximum (FWHM)
of the H$\alpha$ emission
line. Of those, $\Delta m$, $\alpha$ and $W_\lambda$ represent indicators for 
the mass-transfer rate $\dot{M}$. This is based on accretion discs in 
high-$\dot{M}$ systems being bright, blue and optically thick
\citep*[e.g.][]{franketal02-1}, which in turn corresponds to a small 
$\Delta m$, high $\alpha$ and small $W_\lambda$, respectively. It should be 
noted that in $\Delta m$ we do ignore that different bandpasses (may) have 
been used for measuring $m_\mathrm{max}$
and $m_\mathrm{min}$. Considering the typical colour space of CVs 
\citep[e.g.][]{bruch+engel94-1,szkody94-2}, we estimate the respective 
uncertainty to $\sim$0.5 mag. 

However, there are a number of other factors
that affect the above parameters. The brightness of the disc, and thus 
$\Delta m$,
depends on the system inclination \citep{warner95-1}. Regarding
the line strengths, $W_\lambda$ can potentially sample more than one line, as 
is the case for BS Sgr, and furthermore is of course only then an indicator of 
the optical thickness of the disc if (most of) the line actually originates in 
the disc. Finally, the validity of the parameter $\alpha$ that measures the 
slope of the SED depends strongly on the correction for interstellar 
extinction and the determination of the instrumental response function. 
Summarizing the discussion in Section \ref{results_sec}, we point out that
the extinction correction using the tabulated data does not take into
account potential small-scale dust concentrations, nor the actual position
of the target with respect to the dust along the line of sight. In general,
the correction will represent an upper limit, and at least in the case
of V999 Sgr the resulting SED likely suffers from overcorrection. Secondly, the 
instrumental response function relies on the data of a single standard star
that were taken on a different night than the spectroscopy of the targets. 
Finally, the
central CCD chip and the one covering the red part of the spectrum 
(redwards of 5870 {\AA}) likely have a different spectral efficiency that 
could not be corrected for with the available standard star data. This
causes the SED in the `red' chip to have a somewhat steeper slope than 
in the central chip. Therefore, while it should be feasible to 
discuss the values derived here for $\alpha$ within the present 
sample, only a very rough comparison with other works will be possible.
Nevertheless, we point out that the $\alpha$ parameters derived for 
the present post-novae (excluding V999 Sgr) lie well within the previously
observed distribution
\citepalias{tappertetal12-1,tappertetal14-1}. As a consequence of the 
various respective uncertainties, the parameters above cannot be used
as unambiguous indicators for the intrinsic properties of the post-novae on
their own, but have to be evaluated together. In this way, and additionally 
taking into account the FWHM of the emission lines that serves as an 
indicator of the system inclination, it should be possible to obtain at least a 
rough picture of the nature of our post-novae.

We first examine the `blue' and `red' $\alpha$ parameters. As mentioned in 
Section \ref{results_sec}, the red slope is usually steeper than the blue one. 
For HS Sge, BS Sgr and GR Sgr that difference amounts to 
$\sim20\sigma-30~\sigma$. 
In the case of V528 Aql the difference is much more pronounced. Since this
system is an unresolved visual binary (Section \ref{v528aql_sec}), it is 
likely that the spectrum contains a certain level of contamination by the 
neighbour, so that it cannot be described by a power law. In V999 Sgr,
on the other hand, blue and red $\alpha$ are identical within $3~\sigma$.
Here we find by far the largest $\alpha$, even in comparison to previous
work \citepalias{tappertetal12-1,tappertetal14-1}, which lets it appear
likely that this is due to overcorrection of the interstellar extinction. For 
such a steep slope, apparently the differences in the chip-specific 
sensitivities become negligible.

For a quantitative comparison we restrict ourselves to the blue $\alpha$
values, mainly because in V528 Aql we suspect that the red part is affected
by the contribution of the close neighbour (Section \ref{v528aql_sec}). We
find that V528 Aql, HS Sge and GR Sgr have similar values and form the middle 
part of the distribution, while BS Sgr and V999 Sgr represent the lower and
upper end, respectively. The picture in BS Sgr is rather unclear. The slope
parameter $\alpha = 1.19$ suggests a comparatively low $\dot{M}$, which is 
supported by the strengths of the emission lines, although one has to take 
into account that there is a significant contribution of He{\sc ii} to the 
hydrogen lines. On the other hand, the small eruption amplitude indicates a 
bright source
in the system, and this is usually interpreted as the presence of a bright
accretion disc. Both maximum and minimum brightness seem to be known with
sufficient precision so that there is no reason to suspect a significantly
larger $\Delta m$. We will come back to this system below.

As has already been mentioned, the steep slope of the SED in V999 Sgr
points to an overcorrection of the interstellar extinction. This is supported
by a comparison with GR Sgr, which has otherwise very similar parameters 
$W_\lambda$, FWHM, and likely even $\Delta m$, since the estimate
of \citep{duerbeck87-1} for $m_\mathrm{max}$  (see Section \ref{grsgr_sec}) 
suggests an eruption amplitude for GR Sgr of $\Delta m \sim$9 mag.
As evidenced by the small $\Delta m$ and $W_\lambda$, with the blue Balmer
emission lines being in absorption troughs, GR Sgr and V999 Sgr are clearly 
still driving a very high $\dot{M}$, while V528 Aql and HS Sge are operating
at a lower $\dot{M}$. Remarkably, the former two systems are considerably
`older' than the latter two, and with respect to the hibernation scenario
\citep{prialnik+shara86-1,sharaetal86-1} one would expect the opposite
\citep*{kovetzetal88-1}.

In three of the novae, V999 Sgr, HS Sge and BS Sgr
we detect the \mbox{He\sc ii} $\lambda$5412 emission line, which can be
taken as evidence for the presence of a magnetic white dwarf. In BS Sgr
additionally all other \mbox{He\sc ii} corresponding to the transition to
the third excitement level can also be identified, indicating that the
magnetic field in this post-nova is particularly strong. However, the
eruption amplitude is comparatively small 
\citep[e.g.~fig.5.4 in][]{warner95-1}, indicating the presence of an
accretion disc, and thus making it a good candidate for an intermediate
polar. 
In fact, the spectral characteristics are similar to
systems like DQ Her and BT Mon \citep{williams83-1}, the latter of which
also appears to show the same type of decline light curve as BS Sgr
\citep{stropeetal10-1}.
The small FWHM indicates a comparatively low system inclination, 
which also agrees well with the small eruption amplitude.

Our discussion above has assumed that the objects recovered by us are indeed
post-novae, i.e.~that they correspond to the reported nova eruptions. However,
in principle they could also be other CVs that are simply close to the
position of the nova. This concerns in particular the three objects HS Sge,
BS Sgr and V999 Sgr, that were selected via their colour. Still, the space 
density
of CVs is comparatively small. The highest theoretical estimates are based
on models of CV birthrates and are of the order of $10^{-4}~\mathrm{pc}^3$
\citep{dekool92-1,kolb93-1}, while observations still show at most a tenth
of that value \citep[see the summary by][]{pretorius14-1}. Furthermore,
all three objects appear to belong to the high-$\dot{M}$ CVs, which
represent $\le$1\% of the intrinsic CV population \citep{kolb93-1}. It
thus appears rather unlikely that such a CV would be found in the immediate
($\le$5 arcsec) neighbourhood of a nova eruption and not be the post-nova.
A proper calculation of the respective probability would have to include
a model of the Galactic disc structure and take into account interstellar 
extinction (that has already turned out to be problematic even for our
rough analysis), and is beyond the scope of this paper. We can, however,
estimate the probability to recover the original nova in our data. To this
purpose we define the limiting magnitude of our photometry using the 
5$\sigma$ criterion, and thus as the faintest star that has been measured with
a signal-to-noise ratio S/N = 5, translating to a $V$ magnitude uncertainty of
0.198 mag. The resulting limits are 22.4, 23.8 and 22.6 mag, for
the photometries of HS Sge, BS Sgr and V999 Sgr, respectively. Taking the
reported maximum brightness for those objects we would thus have detected
novae with eruption amplitudes of 15.7, 13.8 and 14.4 mag in above
sequence. In Fig.~\ref{amphis_fig} we show the distribution of the
eruption amplitudes $\Delta m$ for all novae that erupted before 1986. This 
limit was chosen to ensure that the contribution of the shell to the optical 
brightness is negligible. We need to emphasize at this point that
$\Delta m$ is calculated as the difference between the maximum brightness
during eruption and the quiescence brightness of the post-nova. This is
important since the latter can be considerably different from the brightness
of the pre-nova \citep{schaefer+collazzi10-1}. The data for 
Fig.~\ref{amphis_fig} have been taken mainly from \citet{downesetal05-1} and
\citet{stropeetal10-1}. In a few cases their quiescent brightnesses differ
significantly ($>$0.5 mag) from data published elsewhere and wherever we felt
that the latter were more convincing we used those. Specifically, this concerns 
V446 Her \citep*{honeycuttetal11-1}, V849 Oph \citep*{shafteretal93-1} and 
CP Pup \citep{patterson+warner98-1}. Additionally, we used the data from
Paper I--IV for the respective objects. The total number of 
novae in this distribution amounts
to 80. In this way we find that 79 objects, and thus 98.75 per cent, have 
$\Delta m < 15.7$ mag. Consequently, the probability to detect the post-nova
HS Sge with the respective photometric data is $\sim$99 per cent. The 
corresponding values for BS Sgr and V999 Sgr are $\sim$94 per cent (75 novae 
with $\Delta m < 13.8$ mag) and 97.5 per cent (78 novae with $\Delta m < 14.4$ 
mag), respectively. Note
that this does not take into account the possibility that a post-nova could 
have atypical colours, and thus, while actually being present in the 
photometric data would not have been recognized as a CV. Still, in each case 
the photometry would have detected a typical post-nova with a high probability.
Taking into account the low space density of CVs, we thus are confident 
that the CVs detected here indeed represent the post-novae.

\section{Summary}
\label{concl_sec}

\begin{enumerate}
\item We have confirmed the five post-nova systems V528 Aql, HS Sge, BS Sgr,
GR Sgr and V999 Sgr spectroscopically. For the three fields of V1301 Aql,
V1151 Sgr and V3964 Sgr, we have identified likely candidates for the 
post-novae.
\item V528 Aql has the appearance of a non-magnetic CV with a comparatively
medium high mass-transfer rate. The emission lines seem sufficiently strong
and broad to suggest follow-up time-resolved spectroscopic data to determine 
the orbital period. Such observations will have to take into account that
the post-nova has a close visual neighbour and thus will need good seeing
conditions.
\item For HS Sge, we find that the H$\alpha$ emission line likely still 
includes some contribution from the ejected material. The spectrum furthermore
contains comparatively strong He{\sc ii} emission, which might indicate
the presence of a magnetic white dwarf. While the object is faint, the width
of the emission line is such that it indicates a medium to high system
inclination. This is thus an attractive target for time series photometry
since the light curve is likely to show orbital features.
\item The spectrum of BS Sgr 
presents a large number of He{\sc ii} emission lines, making it a good 
candidate for a magnetic CV. Since the small eruption amplitude indicates the 
presence of a comparatively bright accretion disc, it appears that BS Sgr is
an intermediate polar. The emission lines are narrow, but still appear to
show radial velocity variations even at our low spectral resolution. This
makes them unlikely to originate in a disc, and perhaps points to a location
on the potentially irradiated secondary star. Nevertheless, while the lines in 
BS Sgr are narrow, they are also strong, and with a quiescent brightness of 
17.9 mag time series spectroscopy should be feasible with a telescope of the 
4\,m class.
\item Finally, both GR Sgr and V999 Sgr are very obviously post-novae that 
still drive a very high $\dot{M}$. The few discernible emission lines are both 
weak and narrow, and thus any attempt to derive the orbital period will need 
high S/N, high spectral resolution data.  While both objects 
are comparatively bright, this still might require one of the larger 
telescopes.
\item It is curious that two of the `oldest' novae in our sample, GR Sgr 
and V999 Sgr, appear to have
the highest mass-transfer rates, contrary to what one would expect from models
of nova evolution. Still, in a sample as small as the present, this easily
can be due to individual particularities, and not necessarily reflect a general
tendency. A study of a much larger sample will be necessary to properly
address this point.
\end{enumerate}

\section*{Acknowledgements}
We thank the referee for helpful comments.

We are indebted to the Gemini and ESO astronomers who performed the service 
observations.

This research was supported by FONDECYT Regular grant 1120338 (CT and NV).
AE acknowledges support by the Spanish Plan Nacional de Astrononom\'{\i}a y 
Astrof\'{\i}sica under grant AYA2011-29517-C03-01. 

We gratefully acknowledge ample use of the SIMBAD data base, 
operated at CDS, Strasbourg, France, and of NASA's Astrophysics Data System 
Bibliographic Services. {\sc iraf} is distributed by the National Optical 
Astronomy Observatories. 

The Guide Star Catalogue-II is a joint project of the Space Telescope
Science Institute and the Osservatorio Astronomico di Torino. Space
Telescope Science Institute is operated by the Association of
Universities for Research in Astronomy, for the National Aeronautics
and Space Administration under contract NAS5-26555. The participation
of the Osservatorio Astronomico di Torino is supported by the Italian
Council for Research in Astronomy. Additional support is provided by
European Southern Observatory, Space Telescope European Coordinating
Facility, the International GEMINI project and the European Space
Agency Astrophysics Division.

Gemini Observatory is operated by the Association of Universities for Research 
in Astronomy, Inc., under a cooperative agreement with the NSF on behalf of the
Gemini partnership: the National Science Foundation (United States), the 
National Research Council (Canada), CONICYT (Chile), the Australian Research 
Council (Australia), Minist\'{e}rio da Ci\^{e}ncia, Tecnologia e 
Inova\c{c}\~{a}o (Brazil) and Ministerio de Ciencia, Tecnolog\'{i}a e
Innovaci\'{o}n Productiva (Argentina).

All data handling and text processing were done on {\sc opensuse} and 
{\sc ubuntu linux} operating systems.

%\bibliographystyle{mn2e}
%\bibliography{./refs}

\appendix

\section{Finding charts}

We present the finding charts for the novae with previously ambiguous or
unknown positions. 

\begin{figure*}
\includegraphics[width=1.8\columnwidth]{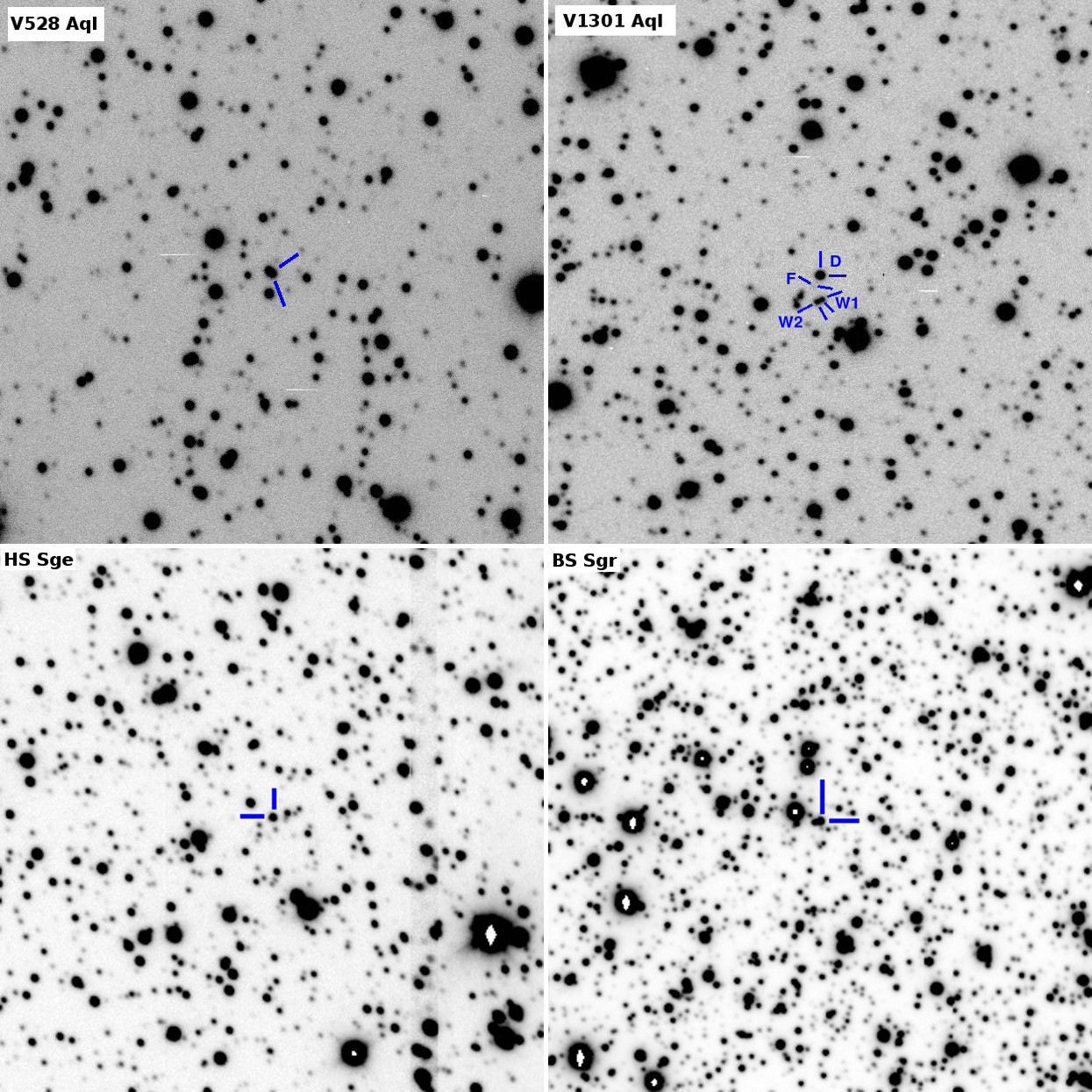}
\caption[]{
Finding charts for V528 Aql, V1301 Aql, HS Sge and BS Sgr, based on images 
taken in $R$ or $r'$ bandpasses. The size of a chart is 1.5 $\times$ 1.5 
arcmin$^2$, and the orientation is such that North is up and East is to the 
left. For details on the labelled objects see the corresponding sections on the 
respective targets. }
\label{fcs1_fig}
\end{figure*}

\begin{figure*}
\includegraphics[width=1.8\columnwidth]{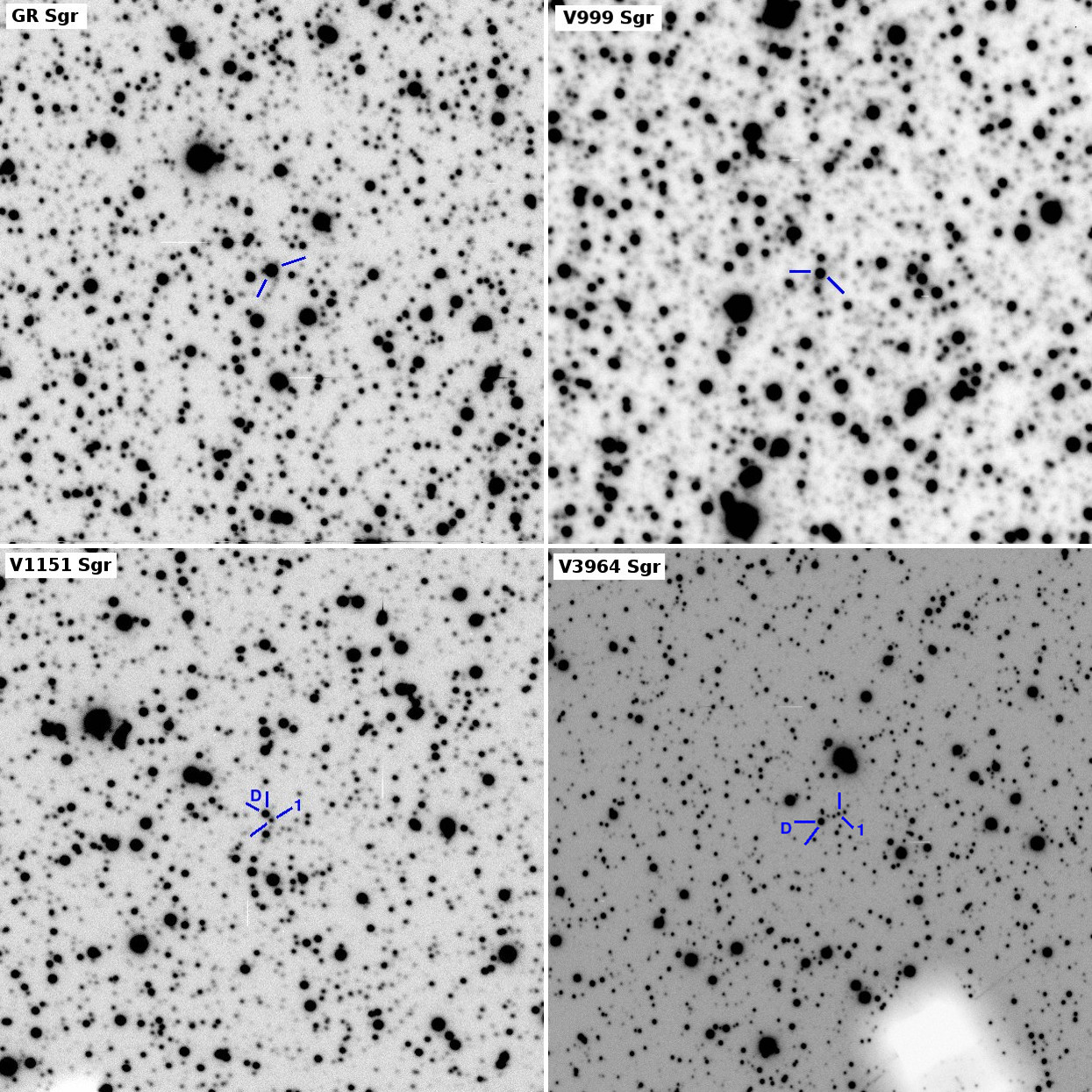}
\caption[]{
Finding charts for GR Sgr, V999 Sgr, V1151 Sgr and V3964 Sgr, based on images
taken in $R$ or $r'$ bandpasses. The size of a chart is 1.5 $\times$ 1.5 
arcmin$^2$, and the orientation is such that North is up and East is to the 
left. 
The white space in the chart of V3964 Sgr is due to vignetting by the
On-Instrument Wavefront Sensor Guiding arm.
For details on the labelled objects see the corresponding sections on the 
respective targets.
}
\label{fcs2_fig}
\end{figure*}

\label{lastpage}

\end{document}